\documentclass[conference]{IEEEtran} % Comment this line out
\IEEEoverridecommandlockouts                              % This command is only
\usepackage{times} % assumes new font selection scheme installed
\usepackage{amsmath} % assumes amsmath package installed
\usepackage{amssymb}  % assumes amsmath package installed
\usepackage{color}
\usepackage{url}

\usepackage[pdftex]{graphicx}
\usepackage{subfig}
\usepackage{multicol}
\usepackage{verbatim}

\usepackage{fancyhdr} 
\usepackage{algorithmic}

\usepackage[font=scriptsize]{caption}

\makeatletter
\newcounter{algorithmbis}
\renewcommand{\thealgorithmbis}{\thesection.\arabic{algorithmbis}}
\def\algorithmbis{\@ifnextchar[{\@algorithmbisa}{\@algorithmbisb}}
\def\@algorithmbisa[#1]{%
  \refstepcounter{algorithmbis}
  \trivlist
  \leftmargin\z@
  \itemindent\z@
  \labelsep\z@
  \item[\parbox{\linewidth}{%
    \hrule
    \hrule
    \noindent\strut\textbf{Algorithm \thealgorithmbis} #1
    \hrule
  }]\hfil\vskip0em%
}
\def\@algorithmbisb{\@algorithmbisa[]}

\makeatother

\usepackage{algorithm} %%for writing algo in pseudo-code
\usepackage{algorithmic}

\title{\LARGE \bf
Learning a set of interrelated tasks by using sequences of motor policies for a strategic intrinsically motivated learner
}

\author{Nicolas Duminy$^{1}$ \and Sao Mai Nguyen$^{2}$ \and Dominique Duhaut$^{1}$ %
\thanks{The research work presented in this paper is partially supported by the EU FP7 grant ECHORD++ KERAAL and by the the European Regional Fund (FEDER) via the VITAAL Contrat Plan Etat Region}%
\thanks{$^{1}$ Nicolas Duminy and Dominique Duhaut are with Universit\'e Bretagne Sud, Lorient, France. {\tt\footnotesize nicolas.duminy@telecom-bretagne.eu} and {\tt\footnotesize dominique.duhaut@univ-ubs.fr}}%
\thanks{$^{2}$ Sao Mai Nguyen is with IMT Atlantique, Lab-STICC, UBL, F-29238 Brest, France. {\tt\footnotesize nguyensmai@gmail.com}
}%
}

\begin{document}

\maketitle
\thispagestyle{empty}
\thispagestyle{fancy}
\lhead{}
\chead{
\texttt{\scriptsize{ N. Duminy, S. M. Nguyen and D. Duhaut, "Learning a Set of Interrelated Tasks by Using Sequences of Motor Policies for a Strategic Intrinsically Motivated Learner," 2018 Second IEEE International Conference on Robotic Computing (IRC), Laguna Hills, CA, US, 2018, pp. 288-291. doi:10.1109/IRC.2018.00061 }}
\vspace{20pt}}
\rhead{}
\cfoot{}

\begin{abstract}
We propose an active learning architecture for robots, capable of organizing its learning process to achieve a field of complex tasks by learning sequences of motor policies, called Intrinsically Motivated Procedure Babbling (IM-PB). %: SAGG-HL (CHANGE NAME OR FIND MEANING).
% Self-Adaptive Goal Generation - Robust Intelligent Adaptive Curiosity (SAGG-RIAC)
% SGIM-PL for the name
The learner can generalize over its experience to continuously learn new tasks. It chooses actively what and how to learn based by empirical measures of its own progress. In this paper, we are considering the learning of a set of interrelated tasks outcomes hierarchically organized.

We introduce a framework called "procedures", which are sequences of policies defined by the combination of previously learned skills . Our algorithmic architecture uses the procedures to autonomously discover how to combine simple skills to achieve complex goals. It actively chooses between 2 strategies of goal-directed exploration: exploration of the policy space or the procedural space. We show on a simulated environment that our new architecture is capable of tackling the learning of complex motor policies, to adapt the complexity of its policies to the task at hand. We also show that our "procedures" framework helps the learner to tackle difficult hierarchical tasks.
\end{abstract}

\section{Introduction}

%Recent efforts in the robotic industry and academic field  to integrate robots in previously human only environments has outlined the need for service robots to continuously learn new tasks. %They would be needed to carry out multiple tasks, especially in open environments, which is still an ongoing challenge in robotic learning.
%Those tasks can be independent and self-contained but they can also be complex and interrelated,  such as combinations of simpler tasks. % to be tackled efficiently.

%The range of tasks those robots need to learn can be wide and even change after the deployment of the robot, we are therefore taking inspiration from the field of developmental psychology to give the robot the ability to learn. 
Taking a developmental robotic approach \cite{DevRobSurvey}, we combine the approaches for active motor skill learning of goal-oriented exploration and strategical learning to learn multiple complex and interrelated tasks. Our algorithm is able to learn a mapping between a continuous space of parametrized tasks (also referred to as outcomes) and a space of parametrized motor policies (sometimes referred to as actions).

\subsection{Active motor skill learning of multiple tasks}

Classical techniques based on Reinforcement Learning \cite{Theodorou2010RAI2IIC} \cite{Stulp2011H} still need an engineer to manually design a reward function for each task. Intrinsic motivation (IM), which triggers curiosity in humans according to developmental psychology \cite{Deci85}, was introduced in highly-redundant robots to make them learn a wider range of tasks, through goal-babbling \cite{Baranes2010} \cite{Rolf10a}.

However, with higher outcome space dimensionalities, their efficiency drops \cite{Baranes2013RAS} due to the curse of dimensionality.%, or when the reachable space of the robot is small compared to its environment. 
%Thus, heuristics such as social guidance can help by driving its exploration towards interesting and reachable space fast.

\begin{comment}

\subsection{Interactive learning}

Combining intrinsically motivated learning and imitation \cite{Nguyen2011PIICDL} has bootstrapped exploration by providing efficient human demonstrations of motor policies and outcomes.

Also, such a learner has been shown to be more efficient if requesting actively a human for help when needed instead of being passive, both from the learner or the teacher perspective \cite{Cakmak2010AMDIT}. This approach is called interactive learning and it enables a learner to benefit from both local exploration and learning from demonstration. Information could be provided to the robot using external reinforcement signals \cite{Thomaz2008CS}, actions \cite{Grollman2010IRS}, advice operators \cite{Argall2008}, or disambiguation among actions \cite{Chernova2009JAIR}. Another advantage of introducing imitation learning techniques is to include non-robotic experts in the learning process \cite{Chernova2009JAIR}. One of the key element of these hybrid approaches is to choose when to request human information or learn in autonomy such as to diminish the teacher's attendance.

\end{comment}

\subsection{Strategic learning}

Approaches where the learner chooses both what (which outcome to focus on) \cite{Baranes2010} and how (which strategy to use) \cite{Baram2004JMLR} to learn are called strategic learning \cite{Lopes2012ICDLE}. They aim at enabling an autonomous learner to self-organize its learning process.

The problem was introduced and studied in \cite{Lopes2012ICDLE}, and implemented for an infinite number of outcomes and policies in continuous spaces by the SGIM-ACTS algorithm \cite{Nguyen2012PJBR}. This algorithm organizes its learning process, by choosing actively both which strategy to use and which outcome to focus on. It relies on the empirical evaluation of its learning progress. It could choose among autonomous exploration driven by IM and low-level imitation of one of the available human teachers to learn more efficiently. It showed its potential to learn on a real high dimensional robot a set of hierarchically organized tasks \cite{duminy_strategic_2016}, so we inspire from it to learn complex motor policies.

\subsection{Learning complex motor policies}

In this article, we tackle the learning of complex motor policies, which we define as sequences of primitive policies.

%We define complex motor policies as combinations of simpler policies. Therefore a complex policy can be described as a sequence of primitive policies.

%A first approach to learning complex motor policies is via-points \cite{Stulp2011H}. They are defined in the robot motor policy space. When increasing the size of the complex policy (by chaining more primitive actions together), we can tackle more complex tasks. However, this increases the difficulty for the learner to tackle simpler tasks reachable using less complex policies. 
We wanted to enable the learner to decide autonomously the complexity of the policy necessary to solve a task, so we discarded via-points \cite{Stulp2011H}. Options \cite{options} are temporally abstract actions built to reach one particular task. They have only been tested for discrete tasks and actions, where a small number of options were used, whereas our new proposed learner is to be able to create an unlimited number of complex policies. 

As we aim at learning a hierarchical set of interrelated complex tasks, our algorithm could use this task hierarchy (as \cite{forestier2016curiosity} did to learn tool use with primitive policies only), and try to reuse previously acquired skills to build more complex ones.  \cite{barto_behavioral_2013} showed that building complex actions made of lower-level actions according to the task hierarchy can bootstrap exploration by reaching interesting outcomes more rapidly.

We adapted SGIM-ACTS to learn complex motor policies of unlimited size. We developed a new mechanism called "procedures" (see Section \ref{sec:procedures}) which proposes to combine known policies according to their outcome. Combining these, we developed a new algorithm called Intrinsically Motivated Procedure Babbling (IM-PB) capable of taking task hierarchy into account to learn a set of complex interrelated tasks using adapted complex policies. We will describe an experiment, on which we have tested our algorithm, and we will present and analyze the results.

%%%%%%%%%%%%%%%%%%% REMOVE IT'S JUST TO SEE SECTION'S LENGTH %%%%%%%%%%%%%%%%%%%%
%\newpage

\section{Our approach}

Inspired by developmental psychology, we propose a strategic learner driven by IM. This learner discovers the task hierarchy and reuses previously learned skills while adapting the complexity of its policy to the complexity.% of the task at hand.

In this section, we formalize our learning problem and explain the principles of IM-PB.

\subsection{Problem formalization}

In our approach, an agent can perform policies $\pi_{\theta}$, parametrized by $\theta \in \Pi$. Those policies induce outcomes  in the environment, parametrized by $\omega \in \Omega$. The agent is then to learn the mapping between $\Pi$ and $\Omega$: it learns to predict the outcome $\omega$ of each policy $\pi_{\theta}$ (the forward model $M$), but more importantly, it learns which policy to choose for reaching any particular outcome (an inverse model $L$). The outcomes $\omega$ are of various dimensionality and  be split in task spaces $\Omega_i \subset \Omega$. 

The policies consist of a succession of primitives (encoded by the same set of parameters $\theta \in \Pi$) that are executed sequentially by the agent. Hence, policies also are of different dimensionality and are split in policy spaces $\Pi_i \subset \Pi$ (where $i$ corresponds to the number of primitives). %Complex policies are represented by concatenating the parameters of each of its primitive policies in execution order. 

\subsection{Procedures\label{sec:procedures}}

As this algorithm tackles the learning of complex hierarchically organized tasks, exploring and exploiting this hierarchy could ease the learning of the more complex tasks. We define procedures as a way to encourage the robot to reuse previously learned skills, and chain them to build more complex ones. More formally, a procedure is built by choosing two previously known outcomes ($t_i, t_j \in \Omega$) and is noted $t_i \boxplus t_j$.

Executing a procedure $t_i \boxplus t_j$ means building the complex policy $\pi_{\theta}$ corresponding to the succession of both policies $\pi_{\theta_i}$ and $\pi_{\theta_j}$ and execute it (where $\pi_{\theta_i}$ and $\pi_{\theta_j}$ reach best $t_i$ and $t_j$ respectively). As $t_i$ and $t_j$ are generally unknown from the learner, the procedure is updated before execution (see Algo. \ref{procedure_refine}) to subtasks $t_1$ and $t_2$ which are feasible by the learner according to its current skill set.

%%It is important to note that even if procedures are task-oriented and take advantage from the task hierarchy, they are only a method to select interesting motor policies to be combined. Furthermore, $t_i$ and $t_j$ are generally unknown from the learner. Hence, $t_i$ and $t_j$ can be updated (see Algo. \ref{procedure_refine}) to subtasks $t_1$ and $t_2$ which are feasible by the learner according to its current skill set. And the complex policy built from the succession of the policies $\pi_{\theta_1}$ and $\pi_{\theta_2}$ (associated with $t_1$ and $t_2$) is then executed by the learner. 

%The algorithm we describe hereafter enables the learner to discover and learn efficient procedures to achieve complex tasks.

\begin{algorithm}[H]
{\footnotesize
    \caption{Procedure modification before execution \label{procedure_refine}}
    \begin{algorithmic}
        \REQUIRE $(t_i, t_j) \in \Omega^2$
        \REQUIRE inverse model $L$
        \STATE $t_1 \gets $Nearest-Neighbour($t_i$)
        \STATE $t_2 \gets $Nearest-Neighbour($t_j$)
        \STATE $\pi_{\theta_1} \gets L(t_1)$
        \STATE $\pi_{\theta_2} \gets L(t_2)$
        \RETURN $\pi_{\theta} = \pi_{\theta_1} \pi_{\theta_2}$
    \end{algorithmic}
}
\end{algorithm}

\begin{comment}
%% ND: Not sure to leave it here or put it in the conclusion
It would be possible, though not studied in this article, to have a recursive process, when building the complex action corresponding to a procedure, in which a subtask could itself be executed by a procedure. In this scenario, the recursive process would end when a motor policy is selected.
\end{comment}

\subsection{Intrinsically Motivated Procedure Babbling}

The IM-PB algorithm (see Algo. \ref{im_pb}) learns by episodes, where an outcome $\omega_g \in \Omega$ to target and an exploration strategy $\sigma$ have been selected.

%The available strategies are: \todo{autonomous exploration of the policy space and autonomous exploration}. You can see the pseudo-code for the SAGG-HL algorithm in Algo. \ref{sgim_hl}.

In an episode under the policy space exploration strategy, the learner tries to optimize the policy $\pi_{\theta}$ to produce $\omega_g$ by choosing between random exploration of policies and local optimization, following the SAGG-RIAC algorithm \cite{Baranes2010} (Goal-Directed Policy Optimization($\omega_g$)).
 Local optimization uses local linear regression.% to interpolate from the known policies reaching an outcome close to $\omega_g$.

In an episode under the procedural space  exploration strategy, the learner builds a procedure $t_i \boxplus t_j$ such as to reproduce the goal outcome $\omega_g$ the best (Goal-Directed Procedure Optimization($\omega_g$)).
It chooses either random exploration of procedures (which builds procedures by generating two subtasks at random) when the goal outcome is far from any previously reached one, or local procedure optimization, which optimizes a procedure using local linear regression. The procedure built is then modified and executed, using Algo. \ref{procedure_refine}.

After each episode, the learner stores the policies and modified procedures executed along with their reached outcomes in its episodic memory. It computes its competence in reaching the goal outcome $\omega_g$ by comparing it with the outcome $\omega$ it actually reached (using normalized Euclidean distance $d(\omega, \omega_g)$). Then it updates its interest model according to the progress $p(\omega_g)$, which is the derivate of the competence, it has made (including the outcome spaces reached but not targeted) in order to choose the strategy and task in the next episode. The interest \textit{interest}($\omega,\sigma)$ of each outcome added depends on both the progress $p(\omega)$ made and the cost $K(\sigma)$ of the strategy used: \textit{interest}($\omega,\sigma) = p(\omega)/ K(\sigma)$. The outcomes reached and the goal are added in their corresponding region, which is then split when exceeding a fixed number of points to discriminate the regions of high and low interest for the learner. The method used is described in \cite{Nguyen2012PJBR}.

\begin{algorithm}[tbh]
{\footnotesize
    \caption{IM-PB \label{im_pb}}
    \begin{algorithmic}
        \REQUIRE the different strategies $\sigma_1,...,\sigma_n$
        \ENSURE partition of outcome spaces $R \gets \bigsqcup_i \lbrace\Omega_i\rbrace$
        \ENSURE episodic memory \textit{Memo} $\gets \varnothing$
        \LOOP
        \STATE $\omega_g, \sigma \gets$ Select Goal Outcome and Strategy($R$)
        \IF{$\sigma$ = Autonomous exploration of procedures strategy}
            \STATE \textit{Memo} $\gets$ Goal-Directed Procedure Optimization($\omega_g$)
        \ELSE
            \STATE $\sigma$ = Autonomous exploration of policies strategy
            \STATE \textit{Memo} $\gets$ Goal-Directed Policy Optimization($\omega_g$)
        \ENDIF
        \STATE Update $L^{-1}$ with collected data \textit{Memo}
        \STATE $R \gets$ Update Outcome and Strategy Interest Mapping($R$,\textit{Memo},$\omega_g$)
        \ENDLOOP
    \end{algorithmic}
    }
\end{algorithm}

The choice of strategy and goal outcome is based on the empirical progress measured in each region $R_n$ of the outcome space $\Omega$, as in \cite{duminy_strategic_2016}.

When the learner computes nearest neighbours to select policies or procedures to optimize (when choosing local optimization in any autonomous exploration strategies and when refining procedures), it actually uses a performance metric (\ref{eq:performance}) which takes into account the cost of the policy chosen:

\begin{equation}
\label{eq:performance}
perf = d(\omega, \omega_g) \gamma^n
\end{equation}

where $d(\omega, \omega_g)$ is the normalized Euclidean distance between the target outcome $\omega_g$ and the outcome $\omega$ reached by the policy, $\gamma$ is a constant and $n$ is equal to the size of the policy (the number of primitives chained).

%%TO DO keep working on algo explanation

%% Maybe add a schema

\begin{comment}

In order to integrate the procedures into the SGIM framework, we need to make some modifications. We created 3 strategies that are using procedures instead of low-level policies to learn new skills.

The random exploration of procedural space creates a random procedure by choosing randomly two outcomes in the outcome space (in the whole outcome space not only among the reached outcomes). 

The autonomous exploration of procedural space performs autonomous exploration of the procedures guided by intrinsic motivation. It performs exactly as the autonomous exploration of actions strategy, except it is using and optimizing procedures instead of actions.

Another strategy called procedural teacher is an interactive strategy. The learner when using this strategy, requests a procedure from the teacher and performs small variations on it.

The new version of SGIM, called SGIM-HL, combines those new procedural strategies with the regular strategies uesd in the SGIM-ACTS version, runs by episodes and chooses actively its target outcome and learning strategy. It can perform autonomous exploration of the policy space (using the SAGG-RIAC algorithm), autonomous exploration of the procedural space, and can also request either policies or procedures from human teachers, respectively imitation teachers and procedural experts.

\end{comment}

%%%%%%%%%%%%%%%%%%% REMOVE IT'S JUST TO SEE SECTION'S LENGTH %%%%%%%%%%%%%%%%%%%%
%\newpage

\section{Experiment}

We designed an experiment with a simulated robotic arm, which can move and interact with objects. It can learn an infinite number of tasks, organized as 6 types of tasks. The robot can perform complex policies of unrestricted size.% (i.e. consisting of any number of primitives), with primitive policies highly redundant and of a high dimensionality.

\begin{comment}

\begin{figure}[t!]
\centering
\includegraphics[width=1\hsize]{fig/setup.jpg}
\caption{Experimental setup: a robotic arm (in black), can interact with the different objects in its environment (a green pen and two red joysticks). The two red box-shape areas show where each joystick can be manipulated). The arm motions are limited by a slightly elastic grey floor.}
\label{fig:setup}
\end{figure}

\end{comment}

%We designed an experiment with a simulated robotic arm of 3 joints that can rotate along the vertical axis. The tip of the arm, referred here as the end-effector, can interact with the robot's environment by taking objects (by hovering close to them). The robot can also change its vertical position to move on a different horizontal plan. A floor is present, which along with limiting the reach of the robot can be drawn upon using a pen which is available among the other objects. Two joysticks also enable to set the position of a video-game character on a screen.

\subsection{Simulation setup}

\begin{figure}[t!]
\centering
\includegraphics[width=1\hsize]{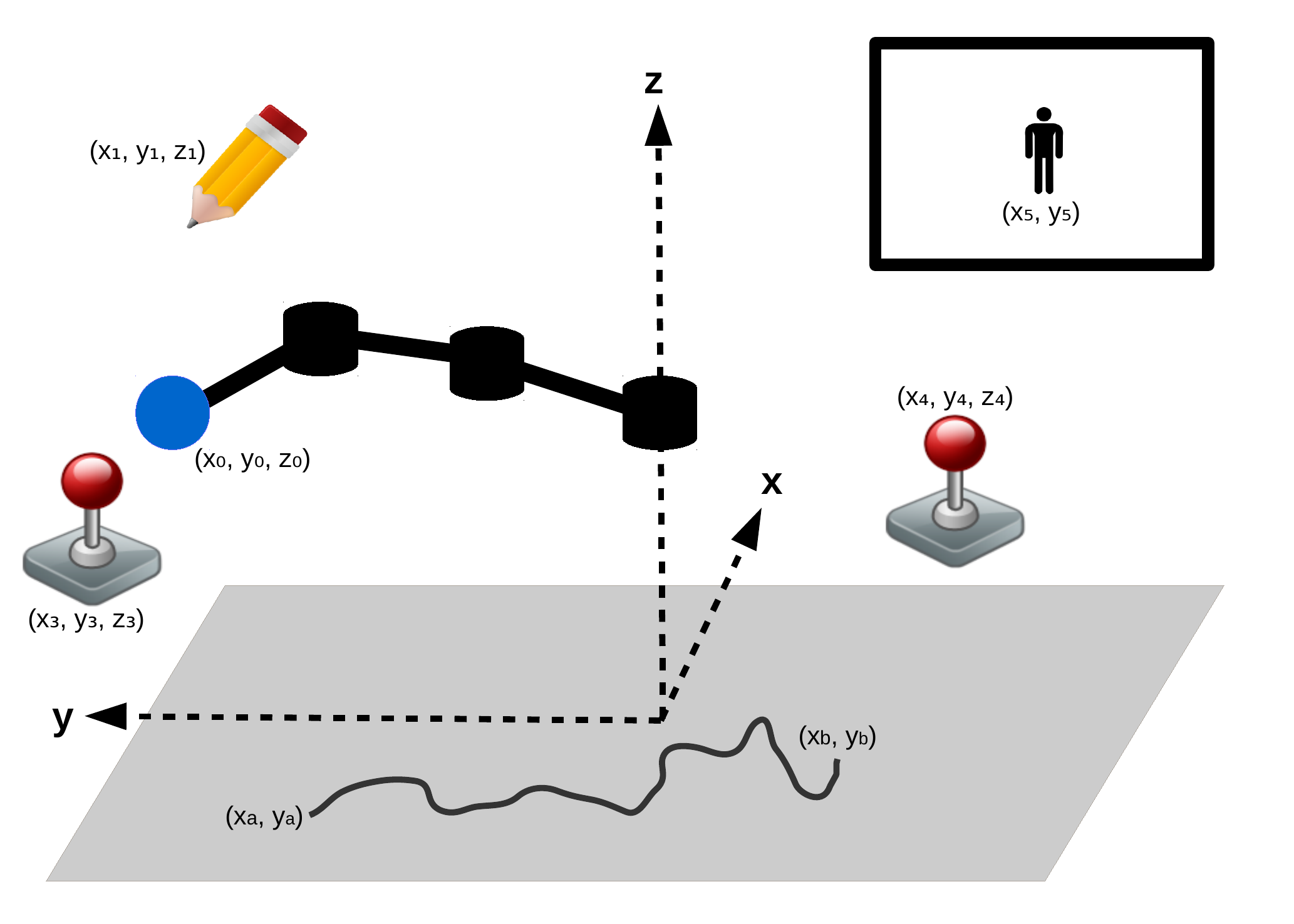}
\caption{Experimental setup: a robotic arm, can interact with the different objects in its environment (a pen and two joysticks). Both joysticks enable to control a video-game character (represented in top-right corner). A grey floor limits its motions and can be drawn upon using the pen (a possible drawing is represented).}
\vspace*{-0.5cm}
\label{fig:setup}
\end{figure}

Fig. \ref{fig:setup} shows the experimental setup (delimited by $(x,y,z) \in \left[ -1;1 \right]^3 $). The learning agent is a planar robotic arm of 3 joints (each link measures $0.33$), with the based anchored in the center of the horizontal plan. It can rotate around the $z$-axis and change its vertical position. The robot can grab objects by hovering its arm tip (blue in Fig. \ref{fig:setup}) close to them, which position is noted $(x_0,y_0,z_0)$. It interacts with:

\begin{itemize}
	\item Pen: position noted $(x_1,y_1,z_1)$, can be moved and draw on the floor, broken if forcing to much on the floor;
    \item Floor which limits motions to $z > -0.2$;
    \item Drawing: last continuous line made when the pen moves on the floor, delimited by first $(x_a, y_a)$ and last $(x_b, y_b)$ point, if the pen is functional;
    \item Joysticks: two joysticks can be moved inside their own cubic-shape volume and control a video-game character, released otherwise, normalized positions respectively at $(x_3,y_3,z_3)$ and $(x_4,y_4,z_4)$;
    \item Video-game character: on a 2D-screen set by the joysticks refreshed only at the end of a primitive policy execution for manipulated joystick, position $(x_5,y_5)$ set by joystick 1 $x$-axis and joystick 2 $y$-axis respectively.
\end{itemize}

%The end-effector of the robot cannot take two objects at the same time without disabling its capacity to grab objects.

The robot can one object at once. Touching an other breaks it, releasing both objects.
It always starts from the same position before executing a policy, and primitives are executed sequentially without getting back to this initial position. Whole complex policies are recorded with their outcomes, but each step of the complex policy execution is recorded as well.

\subsection{Experiment variables}

\subsubsection{Policy spaces}

The motions of each of the three joints of the robot are encoded using a one-dimensional Dynamic Movement Primitive (DMP). We are using the original form of the DMP from \cite{pastor_learning_2009} and we keep the same notations. Each of these one-dimensional DMP $a_i$ (ordered by joint from base to tip) is encoded using its end position $g^{(i)}$, and three basis functions for the forcing term, parametrized by their weights $(\omega_{0}^{(i)}, \omega_{1}^{(i)}, \omega_{2}^{(i)})$. A primitive motor policy is simply the concatenation of those DMP parameters and the fixed vertical position of the arm during the motion $z$:

\begin{align}
\theta &= (a_0, a_1, a_2, z) \\
a_i &= (\omega_{0}^{(i)}, \omega_{1}^{(i)}, \omega_{2}^{(i)}, g^{(i)})
\end{align}

When combining two or more primitive policies $(\pi_{\theta_0}, \pi_{\theta_1},...)$, in a complex policies $\pi_{\theta}$, the parameters $(\theta_0, \theta_1,...)$ are simply concatenated together from the first primitive to the last.

%\begin{equation}
%\theta = (\theta_0, \theta_1,...)
%\end{equation}

\subsubsection{Task spaces}

The task spaces the robot learns are hierarchically organized and defined as: $\Omega_0 = \{(x_0,y_0,z_0)\}$, $\Omega_1 = \{(x_1,y_1,z_1)\}$, $\Omega_2 = \{(x_a, y_a, x_b, y_b)\}$, $\Omega_3 = \{(x_3,y_3,z_3)\}$, $\Omega_4= \{(x_4,y_4,z_4)\}$ and $\Omega_5 = \{(x_5,y_5)\}$.

\begin{comment}
The task spaces the robot learns are hierarchically organized and defined as: the arm tip pose $\Omega_0 = \{(x_0,y_0,z_0)\}$, the still working grabbed pen pose $\Omega_1 = \{(x_1,y_1,z_1)\}$, the last continuous drawing $\Omega_2 = \{(x_a, y_a, x_b, y_b)\}$, the first joystick pose $\Omega_3 = \{(x_3,y_3,z_3)\}$, the second joystick pose $\Omega_4= \{(x_4,y_4,z_4)\}$ and the video-game character pose $\Omega_5 = \{(x_5,y_5)\}$.

\begin{itemize}
    \item $\Omega_0$: the position $(x_0,y_0,z_0)$ of the end effector of the robot in Cartesian coordinates;
    \item $\Omega_1$: the position $(x_1,y_1,z_1)$ of the pen if the pen is grabbed by the robot;
    \item $\Omega_2$: the first $(x_a, y_a)$ and last $(x_b, y_b)$ points of the last drawn continuous line on the floor if the pen is functional $(x_a, y_a, x_b, y_b)$;
    \item $\Omega_3$: the position $(x_3,y_3,z_3)$ of the first joystick if it is grabbed by the robot;
    \item $\Omega_4$: the position $(x_4,y_4,z_4)$ of the second joystick if it is grabbed by the robot;
    \item $\Omega_5$: the position $(x_5,y_5)$ of the video-game character if moved.
\end{itemize}
\end{comment}

%When executing a complex policy, those outcomes will be stored as new skills by the learner at the end of each primitive policy, along with the part of the complex policy executed so far.

\subsection{Evaluation method}

%% Explain testbenches, compared algos
To evaluate our algorithm, we created a benchmark linearly distributed across the  $\Omega_i$, of 27,600 points. The evaluation consists in computing mean Euclidean distance between each of the benchmark outcomes and their nearest neighbour in the learner dataset. %When the learner is incapable to at least reach the outcome space, the evaluation is set to $5$. Then we compute the mean distance to benchmark for each outcome space. The global evaluation is the mean evaluation for the 6 outcome spaces. 
This evaluation is repeated regularly.% distributed timestamps.

%To evaluate our algorithm, we created a benchmark dataset for each outcome space $\Omega_i$, linearly distributed across the outcome space dimensions, for a total of 27,600 points. The evaluation consists in computing the normalized Euclidean distance between each of the benchmark outcome and their nearest neighbour in the learner dataset. When the learner is incapable to at least reach the outcome space, the evaluation is set to $5$. Then we compute the mean distance to benchmark for each outcome space. The global evaluation is the mean evaluation for the 6 outcome spaces. This process is then repeated across the learning process at predefined and regularly distributed timestamps.

Then to asses our algorithm efficiency, we compare its results of algorithms:
RandomPolicy (random exploration of $\Pi$) ,
SAGG-RIAC (exploration of $\Pi$ guided by IM), 
Random-PB (random exploration of policies and procedures),
IM-PB (exploration of the procedural space and the policy space, guided by IM).

% \begin{itemize}
%     \item RandomPolicy: performs random exploration of the policy space $\Pi$;
%     \item SAGG-RIAC: performs autonomous exploration of the policy space $\Pi$ guided by intrinsic motivation;
%     \item Random-PB: performs both random exploration of policies and procedures;
%     \item IM-PB: performs both autonomous exploration of the procedural space and the policy space, guided by intrinsic motivation.
% \end{itemize}

Each algorithm was run 5 times for 25,000 iterations (complex policies executions). The meta parameter was:  $\gamma = 1.2$. %, $p_1 = 0.25$, $p_2 = 0.6$, $p_3 = 0.15$.

%%%%%%%%%%%%%%%%%%% REMOVE IT'S JUST TO SEE SECTION'S LENGTH %%%%%%%%%%%%%%%%%%%%
%\newpage

\section{Results}

\begin{figure}[h!]
\centering
\vspace*{-0.2cm}
\includegraphics[width=0.7\hsize]{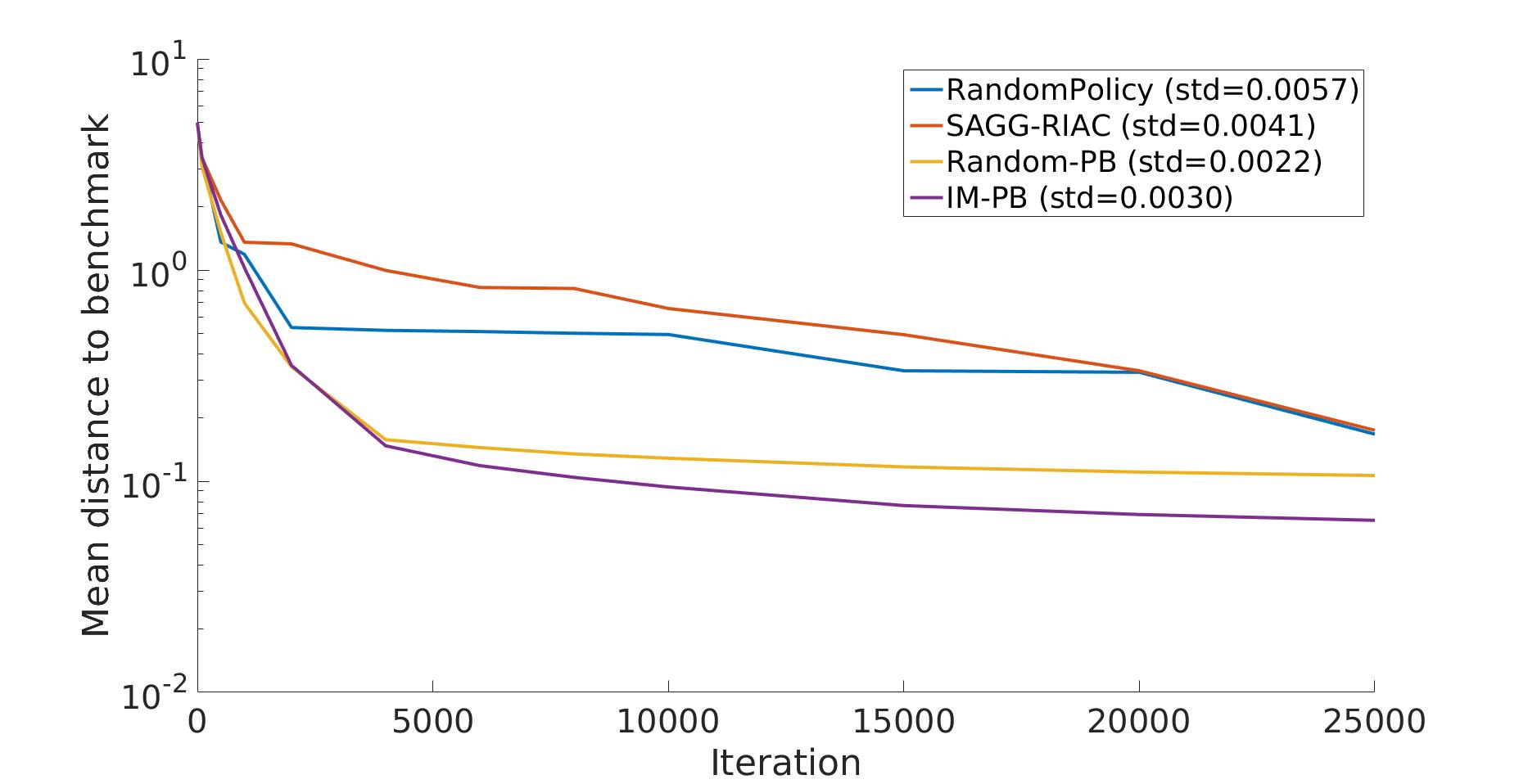}
\caption{Evaluation of all algorithms (standard deviation shown in caption)}
\label{fig:evaluation}
\vspace*{-0.2cm}
\end{figure}

Fig. \ref{fig:evaluation} shows the global evaluation of all tested algorithms, which is the mean error made by each algorithm to reproduce the benchmarks with respect to the number of complete complex policies tried. Random-PB and IM-PB owing to procedures have lower errors than the others even since the beginning. Indeed, they perform better than their downgrades without procedures, RandomPolicy and SAGG-RIAC.
%It is especially true for both autonomous learners, RandomPolicy and SAGG-RIAC, which progress better when using the "procedures" frameworks (corresponding respectively to Random-HL and SAGG-HL).

%We can also see that the SGIM-HL algorithm has a very quick improvement in global evaluation, owing to the bootstrapping effect of the different teachers.

%SAGG-HL goes to 0.065, SGIM-HL reaches it after 8,000 iterations
%Random and SAGG-RIAC go to 0.17 and SGIM-HL reaches it after 500 iterations

%We can also see that the SGIM-HL algorithm has a very quick improvement in global evaluation owing to the bootstrapping effect of the different teachers. It goes lower to the final evaluation of the RandomPolicy and SAGG-RIAC (0.17) after only 500 iterations.

\begin{figure}[h!]
\centering
\includegraphics[width=1\hsize]{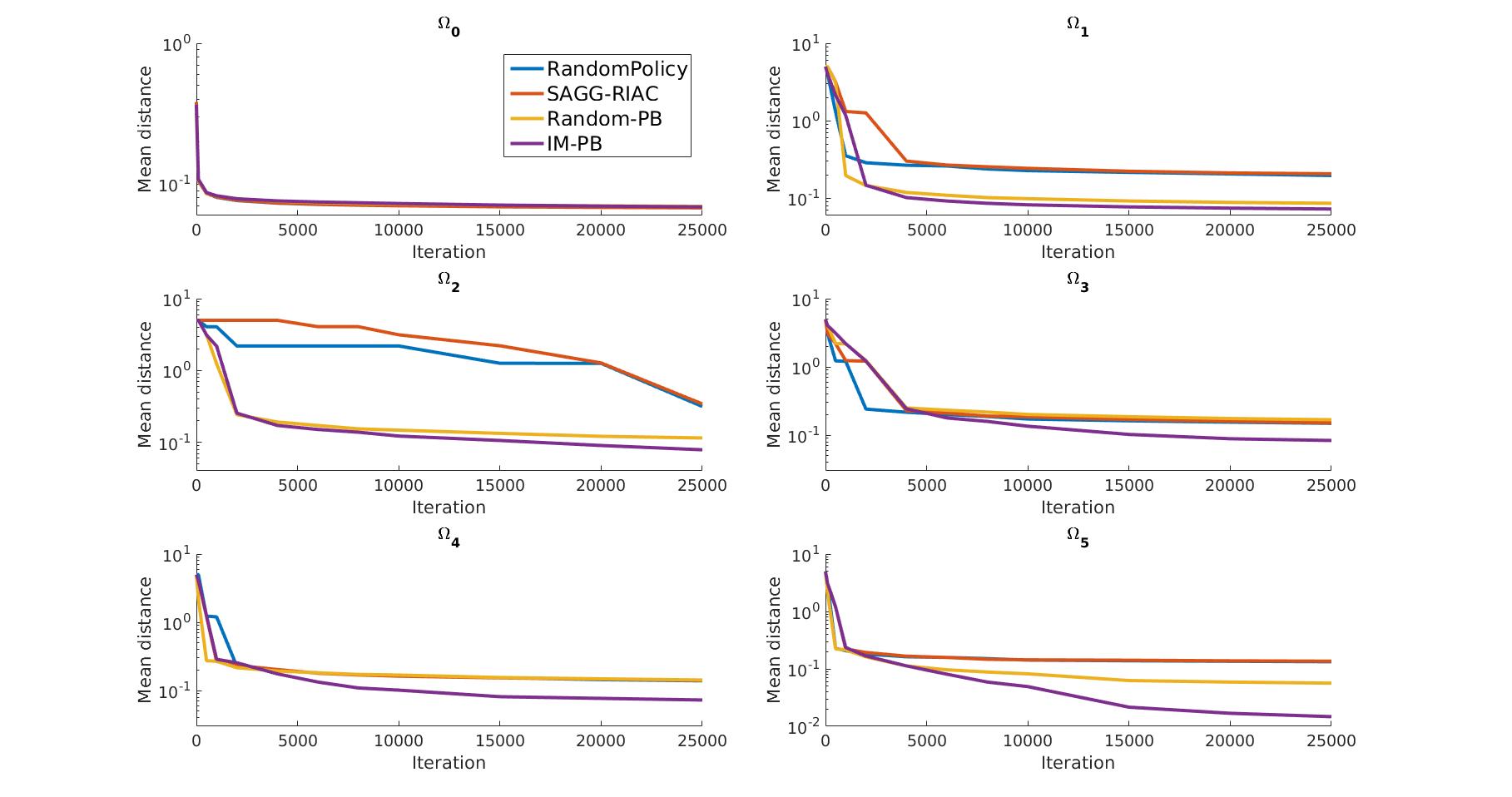}
\caption{Evaluation of all algorithms per outcome space (for $\Omega_0$, all evaluations are superposed)}
\label{fig:evaluation_tasks}
\vspace*{-0.5cm}
\end{figure}

On each individual outcome space (Fig. \ref{fig:evaluation_tasks}), IM-PB outperforms the other algorithms. The comparison of the learners without procedures (RandomPolicy and SAGG-RIAC) with the others shows they learn less on any outcome space but $\Omega_0$ (reachable using single primitives, with no subtask) and especially for $\Omega_1$, $\Omega_2$ and $\Omega_5$ which were the most hierarchical in this setup. So the procedures helped when learning any potentially hierarchical task in this experiment.

\begin{figure}[h!]
\centering
\includegraphics[width=0.7\hsize]{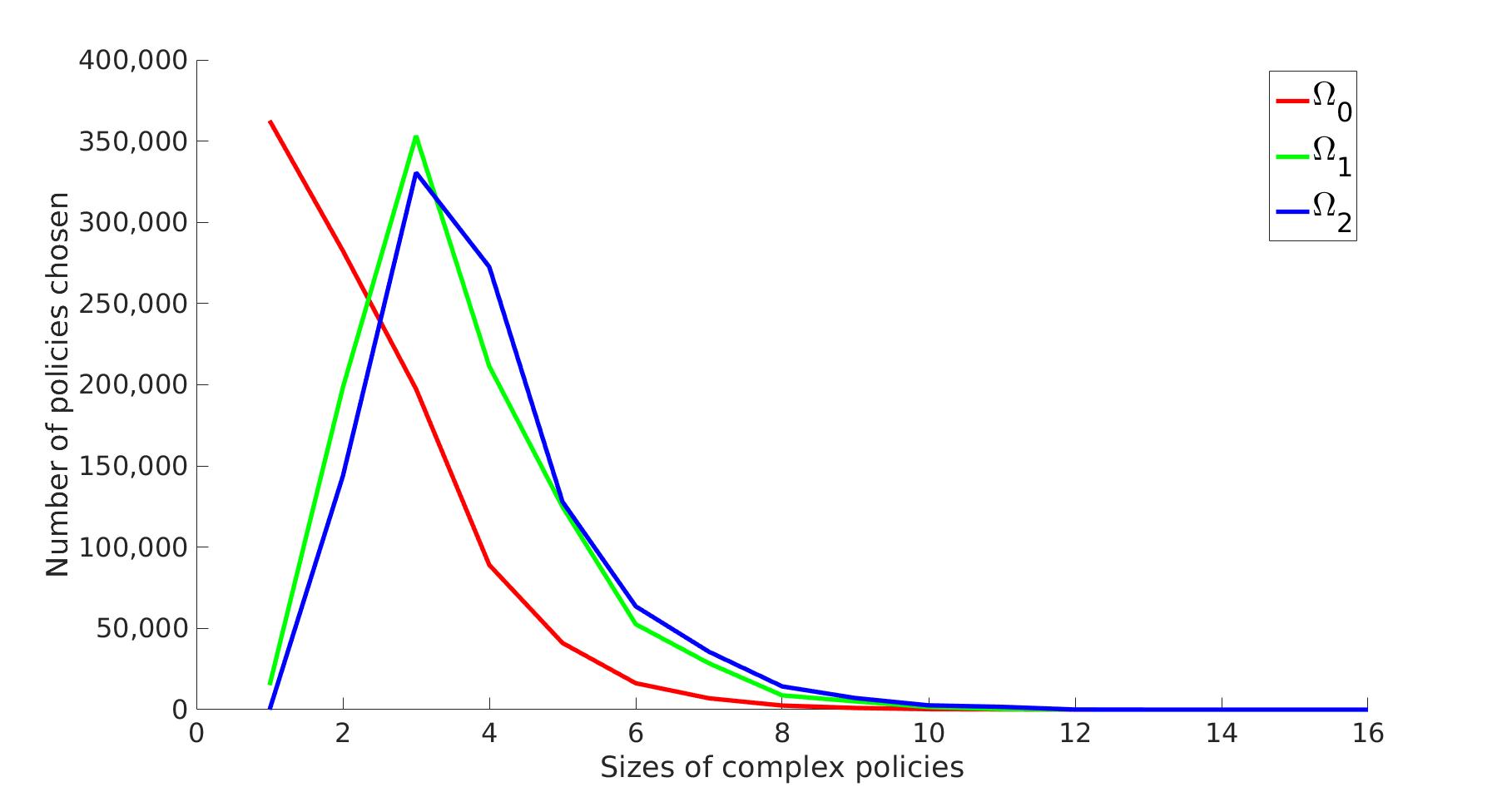}
\caption{Number of policies selected per policy size for three increasingly more complex outcome spaces by the IM-PB learner}
\label{fig:nb_actions}
\end{figure}

We wanted to see if our IM-PB learner adapts the complexity of its policies to the working task. We draw 1,000,000 goal outcomes for each of the $\Omega_0$, $\Omega_1$ and $\Omega_2$ subspaces (chosen because they are increasingly complex) and we let the learner choose the known policy that would reach the closest outcome. Fig. \ref{fig:nb_actions} shows the results of this analysis.

\begin{comment}

\begin{tabular}{| c | c | c | c | c | c | c |}
\hline
 & $\Omega_0$ & $\Omega_1$ & $\Omega_2$ & $\Omega_3$ & $\Omega_4$ & $\Omega_5$ \\
\hline
AutonomousPolicy & & & & & & \\
\hline
AutonomousProcedure & & & & & & \\
\hline
MimicryTeacher1 & & & & & & \\
\hline
MimicryTeacher2 & & & & & & \\
\hline
MimicryTeacher3 & & & & & & \\
\hline
MimicryTeacher4 & & & & & & \\
\hline
MimicryTeacher5 & & & & & & \\
\hline
ProceduralTeacher1 & & & & & & \\
\hline
ProceduralTeacher2 & & & & & & \\
\hline
ProceduralTeacher3 & & & & & & \\
\hline
ProceduralTeacher4 & & & & & & \\
\hline
ProceduralTeacher5 & & & & & & \\
\hline
\end{tabular}

\end{comment}

As we can see on those three interrelated outcome subspaces (Fig. \ref{fig:nb_actions}), the learner is capable to adapt the complexity of its policies to the outcome at hand. It chooses longer policies for  $\Omega_1$ and $\Omega_2$ (size $3$ and $4$ compared to size $1$ for $\Omega_0$). Our learner is capable to correctly limit the complexity of its policies instead of being stuck into always trying longer and longer policies. However, the learner did not increase its policies complexity from $\Omega_1$ to $\Omega_2$, as we hoped.

%As we can see on those three interrelated outcome subspaces (Fig. \ref{fig:nb_actions}), the learner is capable to adapt the complexity of its policies to the outcome at hand. It chooses longer policies for the $\Omega_1$ subspace (policies of size $2$ and $3$ while using mostly policies of size $1$ and $2$ for $\Omega_0$) and even longer for the $\Omega_2$ subspace (using far more policies of size $3$ than for the others). It shows that our learner is capable to correctly limit the complexity of its policies instead of being stuck into always trying longer and longer policies.

%% Show and explain whole evaluation
%% Show and explain task evaluations
%% Compare nb of actions per task spaces

\section{Conclusion and Future work}

With this experiment, we show the capability of IM-PB to tackle the learning of a set of multiple interrelated complex tasks. It successfully uses complex motor policies to learn a wider range of tasks. Though it was not limited in the size of policies it could execute, the learner shows it could adapt the complexity of its policies to the task at hand.

The procedures greatly improved the learning capability of autonomous learners, as we can see by the difference between the Random-PB and IM-PB learners and the RandomPolicy and SAGG-RIAC ones. Our IM-PB shows it is capable to use procedures to exploit both the task hierarchy of this experimental setup and previously learned skills.

However this new framework of procedures could be better exploited, if it could be recursive (defined as a binary tree structure), allowing the refinement process to select lower-levels procedures as one of the policy component. This process could also be used inside the strategical decisions made by the learner when selecting what and how to learn. This strategical choice could also be recursive, allowing the learner to optimize both components of a procedure at once, instead of using the current one-step refinement process.

Also, the procedures are here only combinations of two subtasks, it could be interesting to see if the process can extend to combinations of any number of subtasks.

Finally, proving the potency of our IM-PB learner on a real robotic setup could show its interest for actual robotic application. We are currently designing such an experiment.

\vspace*{-0.08cm}
{\tiny
\bibliographystyle{IEEEtran}
\bibliography{macsi_irc}
}
\end{document}